\documentclass[11pt]{article}
\usepackage[title,toc,titletoc]{appendix}
\usepackage{titlesec,physics}
\usepackage{float,placeins,subcaption}
\usepackage{amsmath,amssymb,latexsym,lscape,amsthm}
\usepackage[table,xcdraw]{xcolor}
\usepackage[final]{graphicx}
\usepackage[colorlinks=true,linkcolor=black,citecolor=black,urlcolor=blue]{hyperref}
\usepackage{inputenc}
\DeclareUnicodeCharacter{0301}{\'{O}}
\usepackage[style=numeric-comp,sorting=none,maxbibnames=9]{biblatex}
\begin{document}

\begin{center}  {\Large \bf Information index augmented eRG to model vaccination behaviour: A case study of COVID-19 in the US}
\end{center}
\smallskip
\begin{center}
	Bruno Buonomo$^1$, Alessandra D'Alise$^{2,3}$, Rossella Della Marca$^1$, Francesco Sannino$^{2,3}$ 
\end{center}

\begin{center} {\small \sl $^1$Department of Mathematics and Applications,
		University of Naples Federico II,\\ via Cintia, I-80126
		Naples, Italy\\ buonomo@unina.it, rdellama@sissa.it}\\
  {\small \sl $^2$Department of Physics, University of Naples Federico II, INFN Section Naples,\\ via Cintia, I-80126 Naples, Italy\\ alessandra.dalise@unina.it, sannino@qtc.sdu.dk}\\
{\small \sl $^3$Quantum  Theory Center ($\hbar$QTC) at IMADA and D-IAS, Southern Denmark University,\\ Campusvej 55, 5230 Odense M, Denmark}
\end{center}

\begin{abstract}
Recent pandemics triggered the development of a number of mathematical models and computational tools apt at curbing the socio-economic impact of these and future pandemics. The need to acquire solid estimates from the  data  led to the introduction of effective approaches such as  the  \emph{epidemiological Renormalization Group} (eRG). A recognized relevant factor impacting the evolution of pandemics is the feedback stemming from individuals' choices. The latter can be taken into account via the  \textit{information index} which accommodates the information--induced perception regarding the status of the disease and the memory of past spread. We, therefore, show how to augment the eRG by means of  the information index.  We first develop  the {\it behavioural}  version of the eRG  and then test it against the US vaccination campaign for COVID-19. 
We find that the behavioural augmented eRG improves the description of the pandemic dynamics of the US divisions for which the epidemic peak occurs after the start of the vaccination campaign. Our results strengthen the relevance of taking into account the human behaviour component when modelling pandemic evolution. To inform public health policies, the model can be readily employed to investigate the socio-epidemiological dynamics, including vaccination campaigns, for other regions of the world.

\end{abstract}


\section{Introduction}
 
Recent pandemics such as the 2009 H1N1 flu and COVID-19  have stressed the need to understand the course of emerging and re-emerging infectious diseases and to provide better supporting tools for public health decision makers  \cite{coburn2009,Covidreview,james21}. This has led to the development of a number of mathematical and computational models able to give useful insights for the interpretation of data, curbing the impact of the pandemics, and in assessing the overall socio-economic impact  \cite{duan2015,vespignani15,MRC,poletti11}. Compartmental models have grown sophisticated  requiring  to determine several parameters from data \cite{merler16,tang2020,trentini2022,SHARMA2021}. In several instances it is hard to obtain reasonable estimates from the available data \cite{RODA2020}. It is for this reason that it is useful to develop effective approaches able to retain salient information while reducing the number of parameters. An example of this type of effective models is the  \emph{epidemiological Renormalization Group} (eRG) framework put forward in \cite{DellaMorte:2020wlc}. It appeared in the wake of the COVID-19 pandemics where it was observed to be an efficient way to encode the overall temporal evolution of epidemic waves. The essential strategy is to devise compartmental models that flow between fixed points  \cite{DellaMorte:2020wlc}.  
The link to the traditional Susceptible-Infected-Recovered (SIR) models was immediately uncovered in \cite{DellaMorte:2020qry} where it was shown that the model can be viewed as a SIR with time-dependent coefficients. The simplest incarnation of the eRG is the Verhulst logistic growth model in which the change in the number of infected is controlled on the right hand side of the differential equation by a product of two zeros in the number of infected, efficiently describing the temporal evolution of a single epidemic wave.  The two fixed points capture the short and long time scaling behaviour of the wave dynamics. The model has been extended to describe different compartments   accounting for the human interaction across different regions of the world (see \cite{cacciapaglia2020} and references therein for a well known application aimed at predicting the COVID-19 second wave pandemic evolution across Europe). Additionally, the eRG approach has been upgraded to  model the vaccination campaigns in US and it has been shown to provide a reasonable understanding of the data \cite{cot2021impact}. Nevertheless, for several regions, the ones for which the vaccination campaign starts before the peak of infected individuals is reached, the eRG overestimates the number of new infected cases. This discrepancy can be attributed to an important aspects, often overlooked in epidemic modelling: the individuals adapt their vaccination behaviours according to the balance between the perceived risk of contagion and the perceived risk of vaccine side effects \cite{TROIANO2021,Lindholt}.

A way to take into account the individuals' choices, including vaccination behaviours, in epidemic models is to employ the  \textit{Behavioural Epidemiology of Infectious Diseases} (BEID) framework \cite{funk2010modelling,manfredi2013modeling,wang2016statistical}. Among the various tools developed within the BEID one is known as the  \textit{information index} which includes two functions representing, respectively, the information--induced perception regarding the status of the disease and the memory of past spread \cite{manfredi2013modeling,wang2016statistical,domasa08,domasaMMNP}. The information index based  approach has shown to be helpful with respect to the non--behavioural models since they can capture important aspects of the epidemic evolution relative, for example, to social distancing and vaccinations   \cite{d2007vaccinating,d2009information,d2022behavioral,KUMAR2019,lopez2022}. Besides the theoretical analysis, such models have been applied to the first COVID-19 wave in Italy as well as to the 2016 Nigerian meningitis outbreak, where an estimate of the information-related parameters has been obtained by using official data released by the local public health system \cite{BBRDMCovid,BBRDMmeningite}.

Here we will augment the eRG by means of the  information index to account for the role of human decision   on the vaccination choices and the consequent impact on the disease dynamics. Inspired by the information--dependent epidemic models \cite{manfredi2013modeling,wang2016statistical}, we  will assume that the choice of individuals to be vaccinated is partially determined on a voluntary basis and depends on the current and past information about the disease. This translates into an information index dependent vaccination rate.  

We therefore first construct the {\it behavioural} improved version of the eRG that we coin {\it BeRG}. The model is then tested against the US vaccination campaign for COVID-19 and the data are used to estimate  a salient information parameter.  We estimate the impact of human behaviour on vaccination compliance via the relative error improvement between the BeRG and the eRG.  

The work is organized as follows: The BeRG is introduced in Section~\ref{sec:model} which is then tested against the US vaccination data in Section~\ref{sec:vaccination}. We offer our conclusions in Section~\ref{sec:conclusions}. 
 
\section{Methodology}
\label{sec:model}

Our main aim here is to build a mathematical model to describe vaccination behaviours during an epidemic outbreak. The model will be designed to account for both human interactions across different geographical areas and individuals' choices within the considered area.
Microscopic models (like lattice, percolation and random walks models) can be used to describe the spread of epidemics in a given geographic area. Computationally simpler compartmental models can be derived as macroscopic analogues \cite{JLCardy_1985}. Among the latter, the eRG framework has been shown to be an effective approach to reproduce real--world epidemic waves while drastically reducing the number of parameters compared to traditional compartmental models \cite{DellaMorte:2020wlc}. 

In the traditional renormalization group approach, widely used in condensed matter and high energy physics, a pivotal role is played by the beta function $\beta$ which governs the time (inverse energy) variation  of the interaction strength among fundamental particles \cite{DellaMorte:2020wlc}. 
In the context of epidemiology, $\beta$ may be interpreted as  the time variation of the epidemic strength $\alpha(t)$, where $\alpha(t)$ is a measure of the interactions among individuals leading to infections from the beginning of the epidemic up to the time $t$  \cite{DellaMorte:2020wlc}. More precisely, $\beta$ and $\alpha$ are linked by the following equation:
\begin{equation}
- \beta (\alpha) = \dfrac{d \alpha}{dt}.
\end{equation}
The quantity $\alpha$ is typically assumed  to be the slowly varying function:
\begin{equation}\label{alphaln}
	\alpha(t)=\ln \mathcal{I}(t),
\end{equation}
where $ \mathcal{I}(t)$ is the cumulative number of infectious individuals, although other monotonic functions of $ \mathcal{I}(t)$ could be in principle used \cite{caccia2022}.

As for  the function $\beta$, its form depends on  the general qualitative behaviour that we expect for an epidemic.
In particular, $\beta$ is assumed to be a polynomial function of $\alpha$ governed by the expected real and/or complex fixed points \cite{DellaMorte:2020wlc,caccia2022} of the model. 

A way to formulate the function $\beta$ is to deduce it from the compartmental model which rules the epidemic dynamics. This is possible if the structure  of the compartmental model allows to derive a first--order differential equation for $\mathcal{I}(t)$.  Then, once specified the function $\alpha$, the expression of  $\beta$  is obtained from 
\begin{equation}
  - \beta (\alpha) = \dfrac{d\alpha}{d\mathcal{I}}\dfrac{d\mathcal{I}}{dt}.
    \label{eq:betap} 
\end{equation}
For example, if we consider a traditional SIR model with time--varying coefficients, then it is possible to derive the differential equation ruling the evolution of $\mathcal{I}$ and it is given by a logistic--like growth model (the detailed procedure is reported in Appendix \ref{App_beta}).\\
Here, we are interested in investigating the impact of information--driven vaccination behaviours on the epidemic strength $\alpha$ given by (\ref{alphaln}). To this aim, we employ the tool of the \textit{information index} in order to build a \textit{behavioural} version of the eRG model. In particular, we want to suitably incorporate such an index in the formulation of the $\beta$ function.\\
In principle, one should
try to derive $\beta$ from suitable behavioural compartmental models where the susceptible individuals are assumed to get vaccinated at an information--dependent rate (see, e.g. \cite{d2007vaccinating}).  However, one can verify that the structure of such kind of models hinders the explicit derivation of  $d{\mathcal{I}}/d{t} $.\\
An elegant and effective way of determining the form of right-hand side of \eqref{eq:betap} is to appeal to the (approximate) temporal symmetries of the problem. For example the following equation\begin{equation}
- \beta (\alpha) = \gamma\alpha  \left( 1 - \dfrac{\alpha}{a} \right),\label{eq:beta0}
\end{equation}
captures the initial and final temporal scale invariance of a single epidemic outbreak. This yields a symmetry-based understanding of the phenomenological Verhulst logistic growth model. Indeed, 
a single wave may be seen as a flow between  two fixed points representing the beginning of the epidemic, where $\alpha=0$  (say, $\mathcal{I}=1$ if (\ref{alphaln}) holds), and the end of the epidemic, when the final size of epidemic strength is reached, i.e. $\alpha=a$. 

The role of the vaccination campaign on the evolution of the epidemic wave is then introduced, following reference \cite{cot2021impact}, by letting $\gamma$ and $a$ to be time--dependent functions satisfying specific differential equations. Since we are interested in the impact of the human behaviour on the vaccination campaign, we assume the time variations of $\gamma$ and $a$ to be affected by the information index $M$.

 In \cite{cot2021impact} we had 
\begin{equation}\label{dgamma}
	\dfrac{d\gamma}{dt}=-c\, \gamma(t_v) \ , \qquad \dfrac{dA}{dt}=-c\, \left(A- e^{\alpha}\right).
\end{equation}
 with $t_v$ the initial time of the vaccination campaign\footnote{The value $t_v$ is estimated via the eRG before taking into account of both the vaccination and its  behavioural impact.}, $A = e^{a}$ and $c$  the fraction of vaccinated individuals in the unit of time. We take into account the impact of the human behaviour by upgrading the original constant $c$ to be a function of the information index $c(M)$.

We further decompose $c(M)$ as: 
\begin{equation}\label{c0+c1}
	c=c_0+c_1(M).
\end{equation}
The first term, $c_0$, represents the fraction of individuals who get vaccinated independently 
of the information (because, e.g., they are strongly in favour of vaccines or belong to a high risk category). The second term,  $c_1(\cdot)$, represents  the fraction of individuals who get vaccinated because of the social alarm caused by the spreading information on the disease.
Specifically, $c_1(\cdot)$ can be modelled as a continuous, differentiable and increasing function of the information index $M$, such that $c_1(0)=0$. 

For $c_1(M)$ we employ the following Holling type II function:
\begin{equation}\label{cH2}
	c_1(M)=c_{1,\max}\dfrac{DM}{1+DM}, 
\end{equation} 
where $c_{1,\max}\geq 0$ is the asymptotic rate of voluntary vaccination when $M\to \infty$, and $D\geq 0$, the \textit{reactivity}, is an adimensional parameter measuring how quickly the individuals react to information. The information index is given by
\begin{equation}
	\label{M}
	M(t)= \int_{-\infty}^t k \, {\cal I}(\tau) H(\tau) \theta e^{-\theta (t-\tau)} d\tau \equiv \int_{-\infty}^t k \,e^{\alpha(\tau)}H(\tau) \theta e^{-\theta (t-\tau)} d\tau.
\end{equation}
The formulation (\ref{M}) derives by the following assumptions: 
\begin{itemize}
    \item [\textit{i})] the choice of individuals of getting vaccinated or not depends on the perceived risk associated to the disease, which is represented by a linear function of the cumulative number of cases  $\mathcal{I}$. This means that, since the analysis of the epidemic is focused on a relatively short time interval, the perceived risk is based on the entire past evolution of the epidemic and not on the current number of new cases. The parameter $k>0$ represents the \textit{information coverage}. The values assumed by $k$ can be thought of as the result of the balance between the disease under–reporting and the media and rumours amplification of the social alarm \cite{d2007vaccinating,manfredi2013modeling,wang2016statistical};
    \item[\textit{ii})] individuals retain memory of the past and this memory is exponentially fading with rate $\theta$. Note that $1/\theta$ is the characteristic memory
length, which can be interpreted as the average time delay in the collection of
the information on the disease \cite{wang2016statistical}. \end{itemize}
In (\ref{M}),  the Heaviside step function $H$ is introduced to restrict the time interval of the relevant information from $(-\infty,t)$ to $(0,t)$ where $t=0$ is the given initial time.

\vspace{.5cm}

 As a consequence of the assumption (\textit{ii}), we can apply the \textit{linear chain trick} \cite{smith2011}, and obtain a differential equation for $M$: 
${dM}/{dt}=\theta (ke^\alpha-M).$
Then, by taking $A = e^{a}$,  the complete model reads:
   \begin{equation}\label{sistema}
	\begin{aligned}
		\dfrac{d\alpha}{dt}&=\gamma \alpha \left(1-\dfrac{\alpha}{\log A}\right)\\
		\dfrac{d\gamma}{dt}&=-c(M)\gamma(t_v)\\
		\dfrac{dA}{dt}&=-c(M)\left(A- e^{\alpha}\right)\\
		 \dfrac{dM}{dt}&=\theta (ke^\alpha-M).
	\end{aligned}
\end{equation}

The discussion above is tailored to describe the spread of infection within a given isolated region of the world. However, in any realistic description of a pandemics say, for example, at the level of US, it is crucial to account for the human exchange across different regions. As summarized in \cite{cacciapaglia2020}, at the eRG level the epidemic diffusion can be efficiently described  by generalizing  Eq.\eqref{sistema} to a number of $W$ coupled differential equations for each region with the  addition of the following interaction term: 
\begin{equation}
\sum_{j\neq i}^W \dfrac{K_{ij}}{n_{mi}}\left(e^{\alpha_j-\alpha_i}-1\right),
\label{eq:deltaI}
\end{equation}
where $K_{ij}\in\mathbb{R}$ is a measure for the number of travellers between the populations $i$ and $j$. The resulting coupled system of differential equations can be thought of flow equations, in the spirit of the Wilsonian renormalisation, with the second term representing a coupling between the different regions. Following \cite{cacciapaglia2020},  it is natural to assume that that the number of travellers is symmetric, i.e. there is no net flow of people between any two regions. This is a reasonable approximation during a short time as immigration only involves a smaller fraction of inhabitants than that involved in the epidemic. Additionally, we suppose that the rate of infected cases within the travelling subset of people is the same as the rate of infected cases for the total population of each region.  We therefore arrive at the following system of equations: 
   \begin{equation}\label{sistemalogvoli}
	\begin{aligned}
		\dfrac{d\alpha _i}{dt}&=\gamma_i \alpha_i \left(1-\dfrac{\alpha_i}{\log A_i}\right)+\sum_{j\neq i}^W \dfrac{K_{ij}}{n_{mi}}\left(e^{\alpha_j-\alpha_i}-1\right)\\
		\dfrac{d\gamma_i}{dt}&=-c(M_i)\gamma_i(t_v)\\
		\dfrac{d A_i}{dt}&=-c(M_i)\left(A_i-e^{\alpha_i}\right)\\
		 \dfrac{dM_i}{dt}&=\theta_i (k_i e^{\alpha_i}-M_i).
	\end{aligned}
\end{equation}
The above defines the Behavioural epidemiological Renormalisation Group (BeRG) including the vaccination impact that can be used for any pandemic across the world. We are now ready to test the model by taking as an example the early stage of the US vaccination campaign for SARS-COV-2. 

\section{The US vaccination campaign as BeRG test-bed}
\label{sec:vaccination}

\noindent {\it (a) Setting up US division and Region-X}

\vspace{0.5cm}

We follow  reference \cite{cot2021impact} and consider US territory in 9 divisions where for geographical reasons we move Maryland and Delaware from the South Atlantic to Mid-Atlantic division as summarised in Table \ref{tab:divisions}. 
The baseline values of the parameters are the ones established in    \cite{cot2021impact}, which refer to the US second wave pandemic period assumed to start around the end of July 2020, as reported in Table \ref{tab:parameters}.

\begin{table}[t!]
    \centering
    \resizebox{\textwidth}{!}{\begin{tabular}{|l|l|l|}
\hline  \rowcolor{lightgray}\multicolumn{3}{|l|}{ \textbf{Division composition} } \\
\hline \rowcolor{lightgray} \textbf{Division name} & \textbf{Code} & \textbf{States within the division} \\
\hline New England & NE & Massachusetts, Connecticut, New Hampshire, Maine, Rhode Island and Vermont \\
\hline Mid-Atlantic & MA & New York, Pennsylvania, New Jersey, Maryland and Delaware \\
\hline South Atlantic & SA & Florida, Georgia, North Carolina, Virginia, South Carolina and West Virginia \\
\hline East South Central & ESC & Tennessee, Alabama, Kentucky and Mississippi \\
\hline West South Central & WSC & Texas, Louisiana, Oklahoma and Arkansas \\
\hline East North Central & ENC & Illinois, Ohio, Michigan, Indiana and Wisconsin \\
\hline West North Central & WNC & Missouri, Minnesota, Iowa, Kansas, Nebraska, South Dakota and North Dakota \\
\hline Mountains & M & Arizona, Colorado, Utah, Nevada, New Mexico, Idaho, Montana and Wyoming \\
\hline Pacific & P & California, Washington, Oregon, Hawaii and Alaska \\
\hline
\end{tabular}}
\caption{The US divisions used in this work. Here, for geographical reasons Maryland and Delaware are moved from South Atlantic
to Mid-Atlantic.}
    \label{tab:divisions}
\end{table}

\begin{table}[t!]
    \centering
    \begin{tabular}{|l|l|l|l|}
\hline \rowcolor{lightgray} \multicolumn{4}{|c|}{ \textbf{Second wave parameters}} \\
\hline \rowcolor{lightgray} \textbf{Division} & \textbf{Code} & $\mathbf{\log{A}}$ & $\mathbf{\gamma}$ \\
\hline New England & NE & 11.006 & 0.214 \\
\hline Mid-Atlantic & MA  & 10.882 & 0.206 \\
\hline South Atlantic & SA  & 10.885 & 0.185 \\
\hline East South Central & ESC & 11.201 & 0.207 \\
\hline West South Central & WSC  & 10.713 & 0.213 \\
\hline East North Central & ENC  & 11.074 & 0.250 \\
\hline West North Central & WNC  & 11.060 & 0.263 \\
\hline Mountains & M  & 11.089 & 0.213 \\
\hline Pacific & P  & 11.535 & 0.171 \\
\hline
\end{tabular}
    \caption{Parameters of the eRG model for the second wave in the 9 divisions and their baseline values, chosen to reproduce
the data until December 16, 2020}
    \label{tab:parameters}
\end{table}

 The divisions assumed to be responsible for the start of the second wave pandemic in US were the ones that experienced a surge in COVID-19 cases during the months of July and August 2020. Following \cite{cot2021impact} we therefore define as \emph{Region-X} the average sum of all the divisions that had a pandemic peak in that period.  \emph{Region-X} effectively defines the initial conditions for the spreading of the pandemic for the coupled set of differential equations.  The baseline values of the parameters are chosen to replicate the number of cases in the combined seven relevant divisions (SA, ESC, NSC, ENC, WNC, M, and P) normalized by the total population. This is achieved by using the parameters $\log A_i$ and $\gamma_i$ for each division to match the  data used in \cite{cot2021impact} available up to December 16, 2020 (Table \ref{tab:parameters}) including the flight data  utilized to determine the coupling matrix entries of $K_{ij}$ (see Table 2 of \cite{cot2021impact}). We focus on the interaction among divisions and Region-X. We first utilize the data on vaccinated individuals to extract the relevant information coverage parameter for each region and then compare the eRG with the BeRG model predictions with the data stemming from both the  partially vaccinated  and new infected  individuals.


\vspace{0.5cm}

\noindent {\it (b) Information coverage via vaccinated individuals}

\vspace{0.5cm}

In this section we compare the model solutions with the data stemming from the first dose vaccinated individuals. This will allow us to estimate the information coverage parameter $k$. To do so, for each census division, we first establish that, in equation \eqref{c0+c1}, $c_{0}=0.0064$ weeks$^{-1}$ since 0.64$\%$ of the total population was vaccinated as of December 28, 2020 \cite{CDCvaccine}. The value of the rate $c_{1,max}$, in equation \eqref{cH2}, is obtained assuming that $c_0+c_{1,max}$ corresponds to the maximum vaccination rate reached during the COVID--19 pandemic in US: it occurred in mid--April 2021 with about 3.35 million daily doses administered in a population of 332 million inhabitants \cite{CDCvaccine}. Namely,  $c_0+c_{1,max}=0.0706$ weeks$^{-1}$, which yields $c_{1,max} =10\,c_0$. 

\begin{figure}[t!]
    \centering
    \includegraphics[scale=0.30]{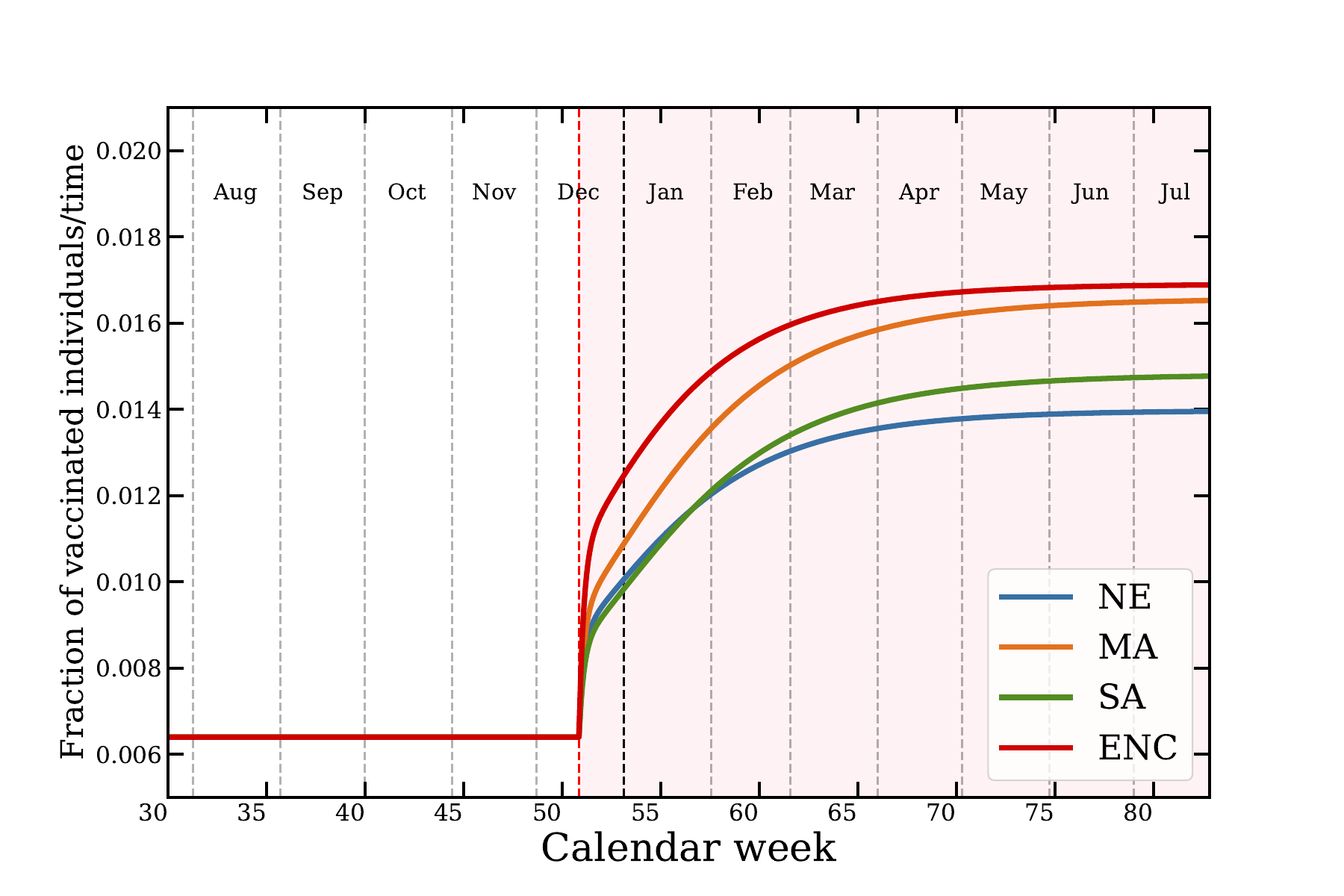}
    \caption{Plot of the fraction of vaccinated individuals per unit of time $c(M(t))$ for the US divisions NE, MA, SA, ESC. }
    \label{fig:cMtotale}
\end{figure}

Equipped with the parameters fixed so far, we estimate the remaining ones by comparing numerical solutions of the system \eqref{sistemalogvoli} with the data from each US division. The preliminary simulations, not reported here show that the $\theta$ parameter does not  influence the results. For this reason, taking into account that in a previous study on the first wave of COVID-19 in Italy it was estimated an average time delay of 3 days in the collection of information on the disease \cite{BBRDMCovid}, we take here the average value in the interval 0--3 days, yielding approximately  $\theta=4$ weeks$^{-1}$. As for the  reactivity  parameter $D$, it modulates the temporal dynamics of the fraction of vaccinated individuals in the unit of time, $c(M)$.  We assume that  $c(M)$ has a steep growth at the beginning of the vaccination campaign, and then it smooths over time  until it reaches a plateau by the end of July 2021 (see Figure \ref{fig:cMtotale}). This scenario is achieved by setting $D=5\times 10^{-6}$.  Similar values for the reactivity parameter $D$ were found in \cite{BBRDMCovid,BBRDMADOMG} when describing  COVID--19  behavioural dynamics. 

\begin{table}[t!]
\centering
\begin{tabular}{| c | c |}
\hline
\rowcolor{lightgray} \multicolumn{2}{|c|}{\textbf{Optimized value of $\mathbf{k}$ and its error}}          \\       
\hline
\rowcolor{lightgray}      \textbf{Division code}       & $\mathbf{k}$   \\ \cline{1-2} 
NE  & $0.70\pm 0.03$              
\\ \hline
MA  & $0.85\pm 0.03$              
\\ \hline
SA  & $0.91\pm 0.03$             
\\ \hline
ESC & $0.47\pm 0.03$            
\\ \hline
WSC & $0.73\pm 0.03$             
\\ \hline
ENC & $0.51\pm 0.03$              
\\ \hline
WNC & $0.51\pm 0.03$           
\\ \hline
M   & $0.61\pm 0.03$            
\\ \hline
P   & $0.74\pm 0.03$          
\\ 
   \hline
\end{tabular}
\caption{Numerical values  of the information coverage $k$ and its error for each US division obtained as explained in the text.}
\label{tab:1}
\end{table}

\begin{figure}[t!]
    \centering
    \includegraphics[scale=0.25]{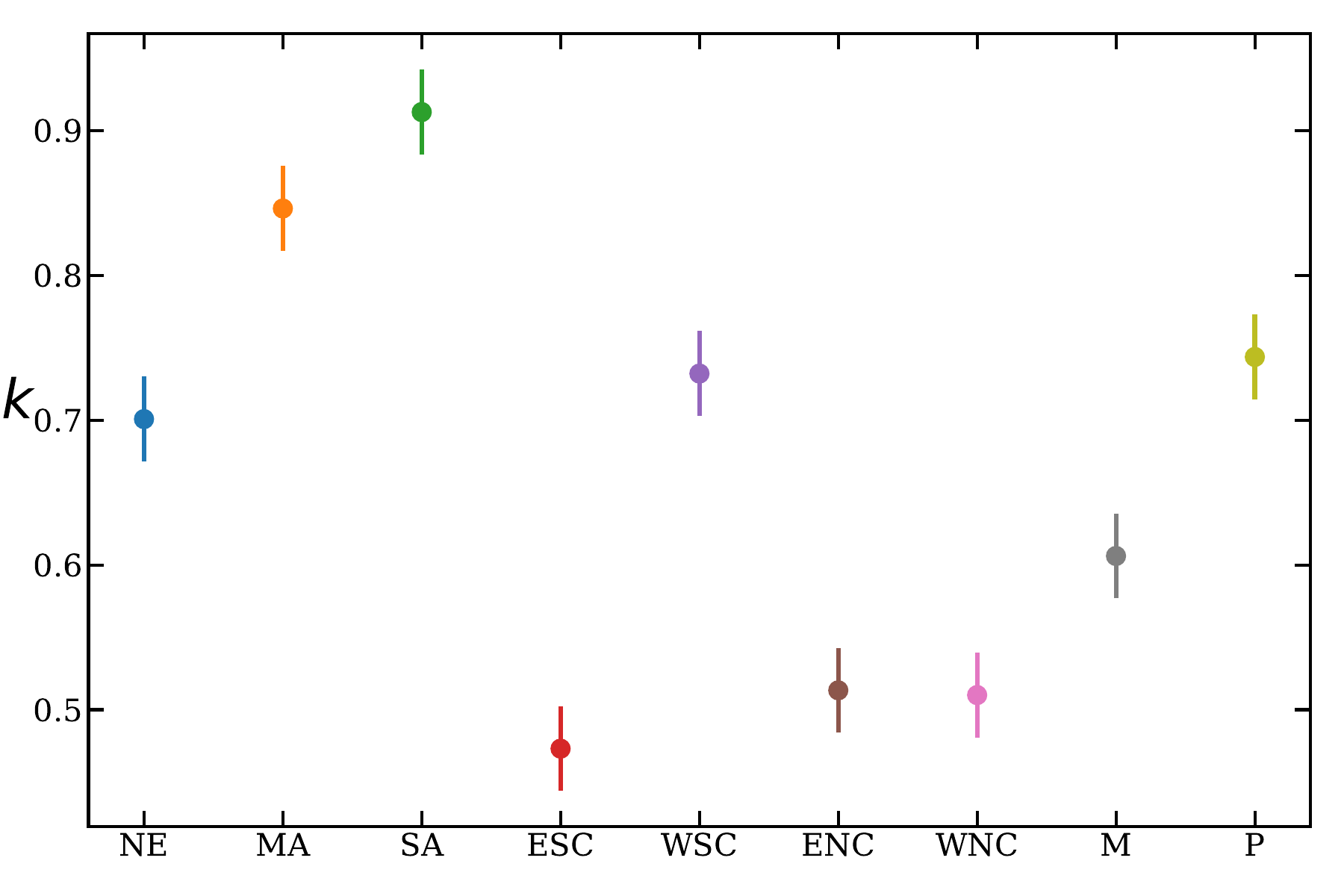}
    \caption{Plot of the optimized values for the information coverage parameter $k$ and its $1\sigma$ error for each US division. }
    \label{fig:listak}
\end{figure}

\begin{figure}[t!]
      \centering
	   \begin{subfigure}{0.30\linewidth}
		\includegraphics[width=\linewidth]{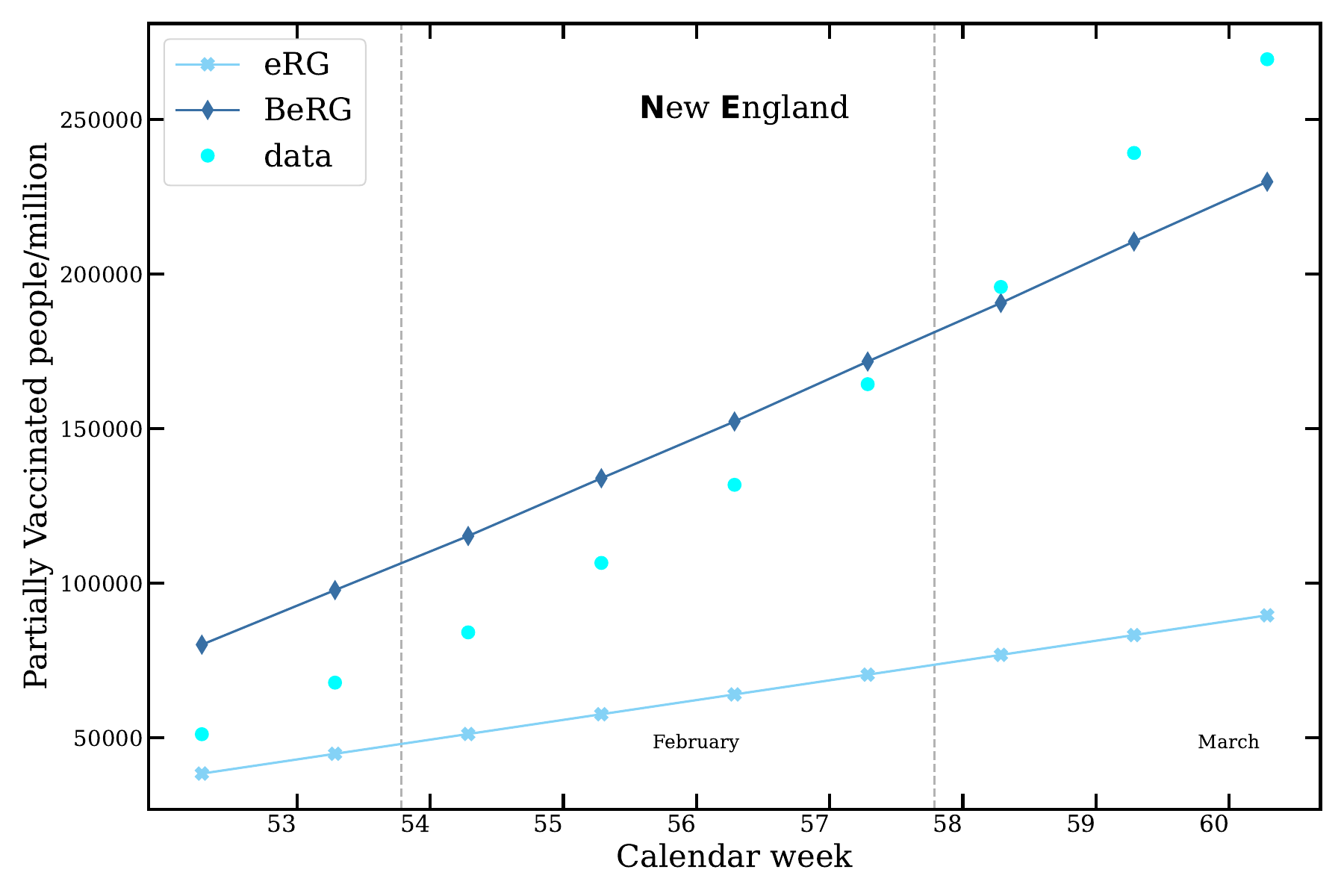}
		\caption{NE}
		\label{fig:NE}
	   \end{subfigure}
	   \begin{subfigure}{0.30\linewidth}
		\includegraphics[width=\linewidth]{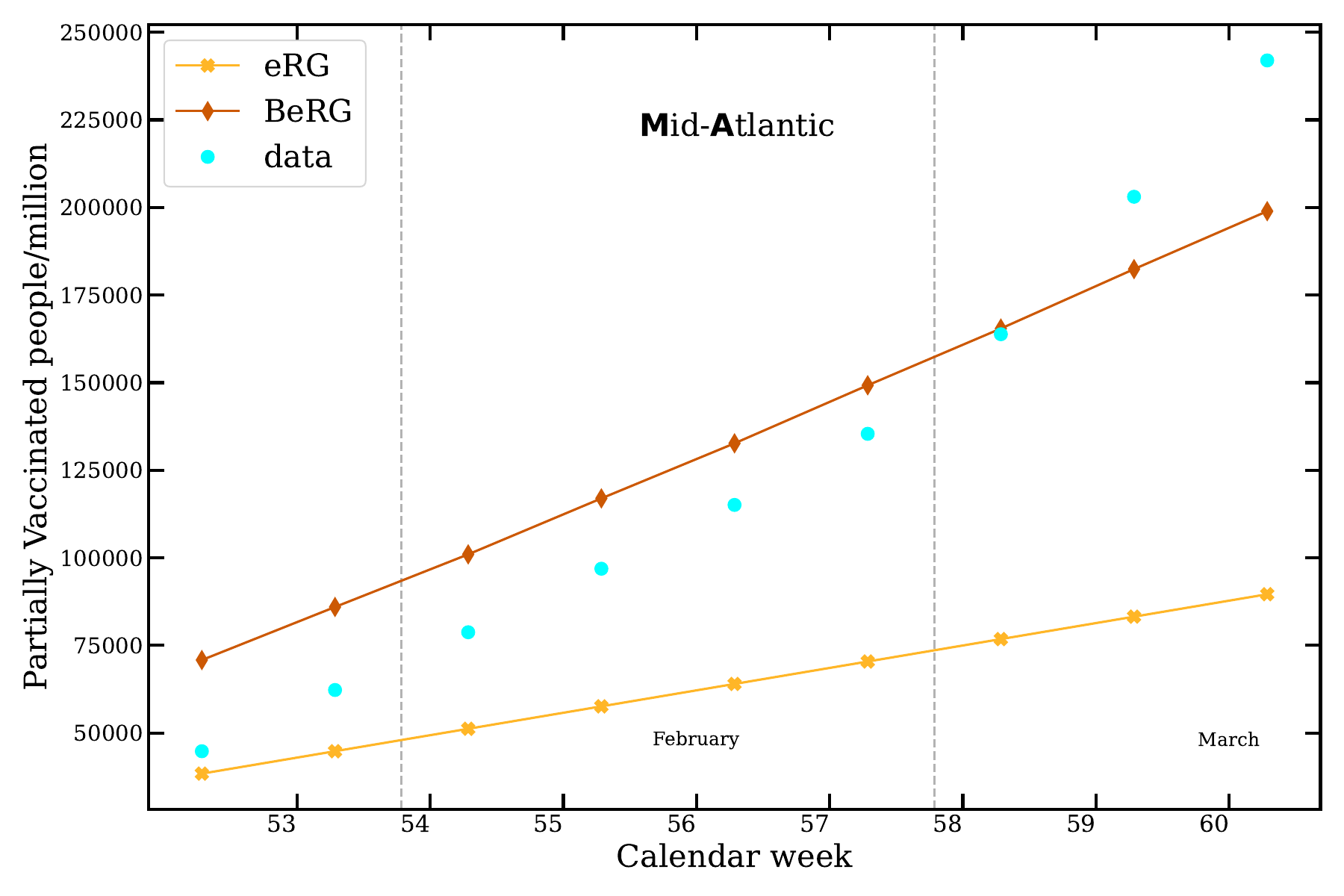}
		\caption{MA}
		\label{fig:MA}
	    \end{subfigure}
     \begin{subfigure}{0.30\linewidth}
		 \includegraphics[width=\linewidth]{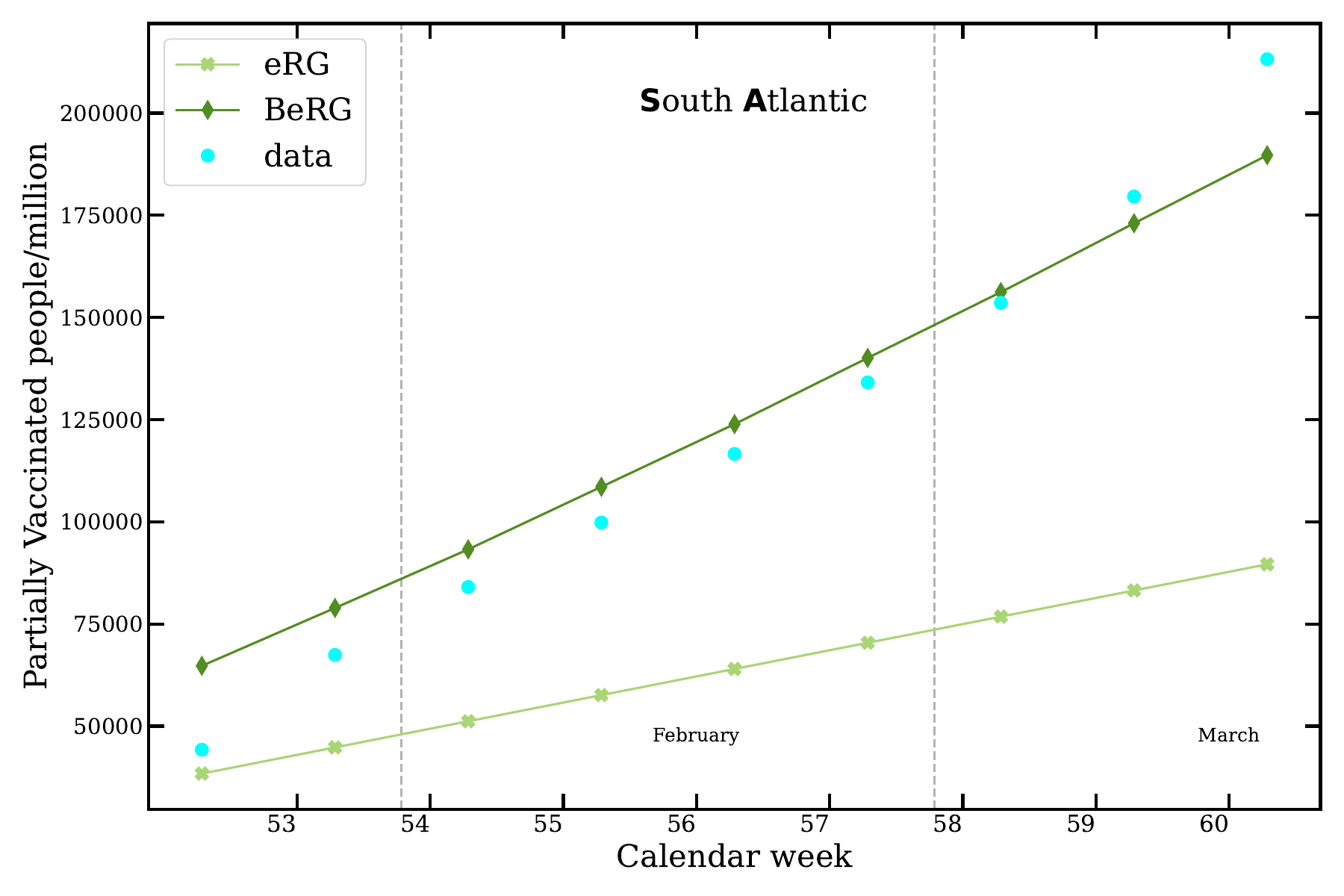}
		 \caption{SA}
		 \label{fig:SA}
	      \end{subfigure}
        \vfill
        \medskip
	       \begin{subfigure}{0.30\linewidth}
		  \includegraphics[width=\linewidth]{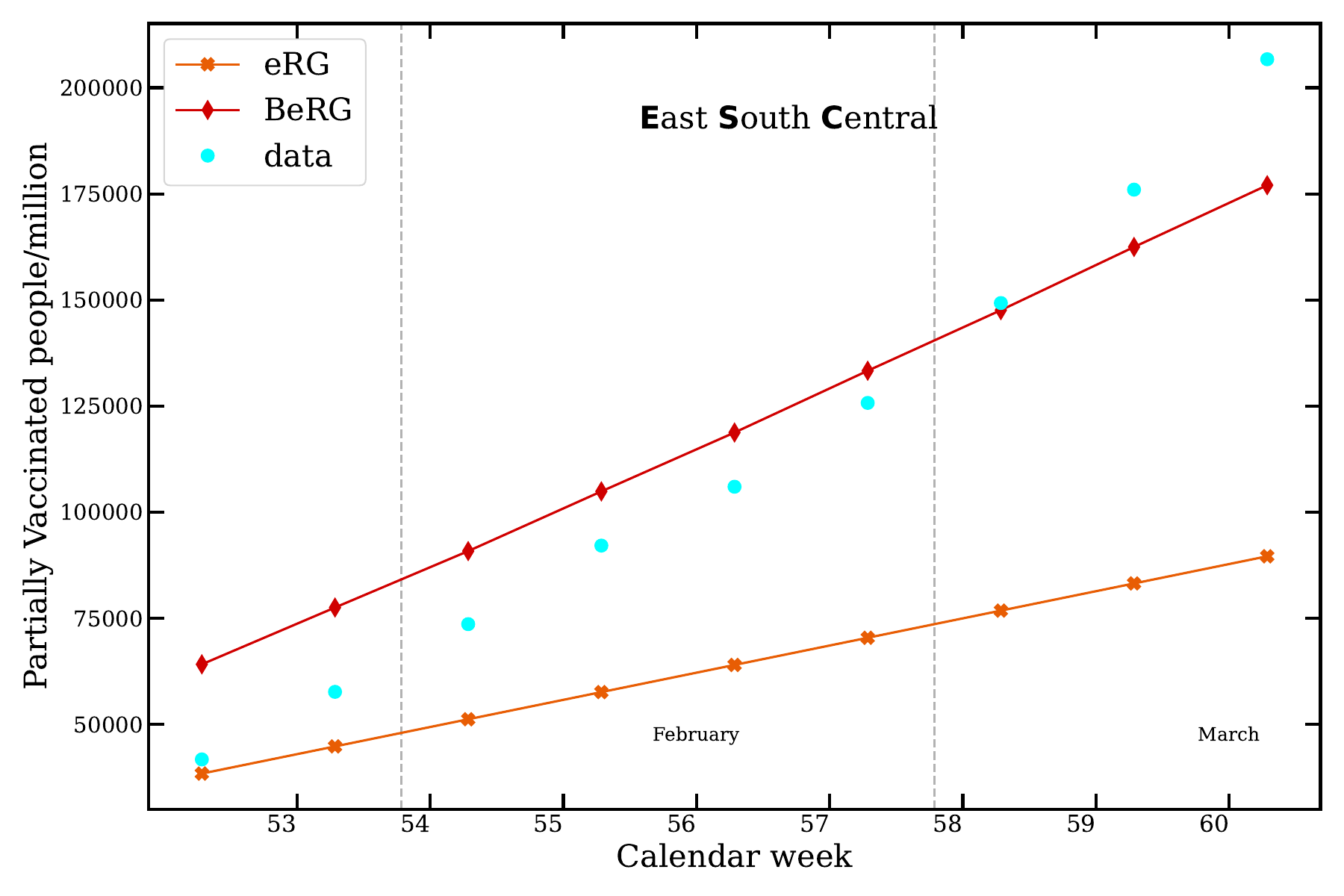}
		  \caption{ESC}
		  \label{fig:ESC}
	       \end{subfigure}
	     \begin{subfigure}{0.30\linewidth}
		 \includegraphics[width=\linewidth]{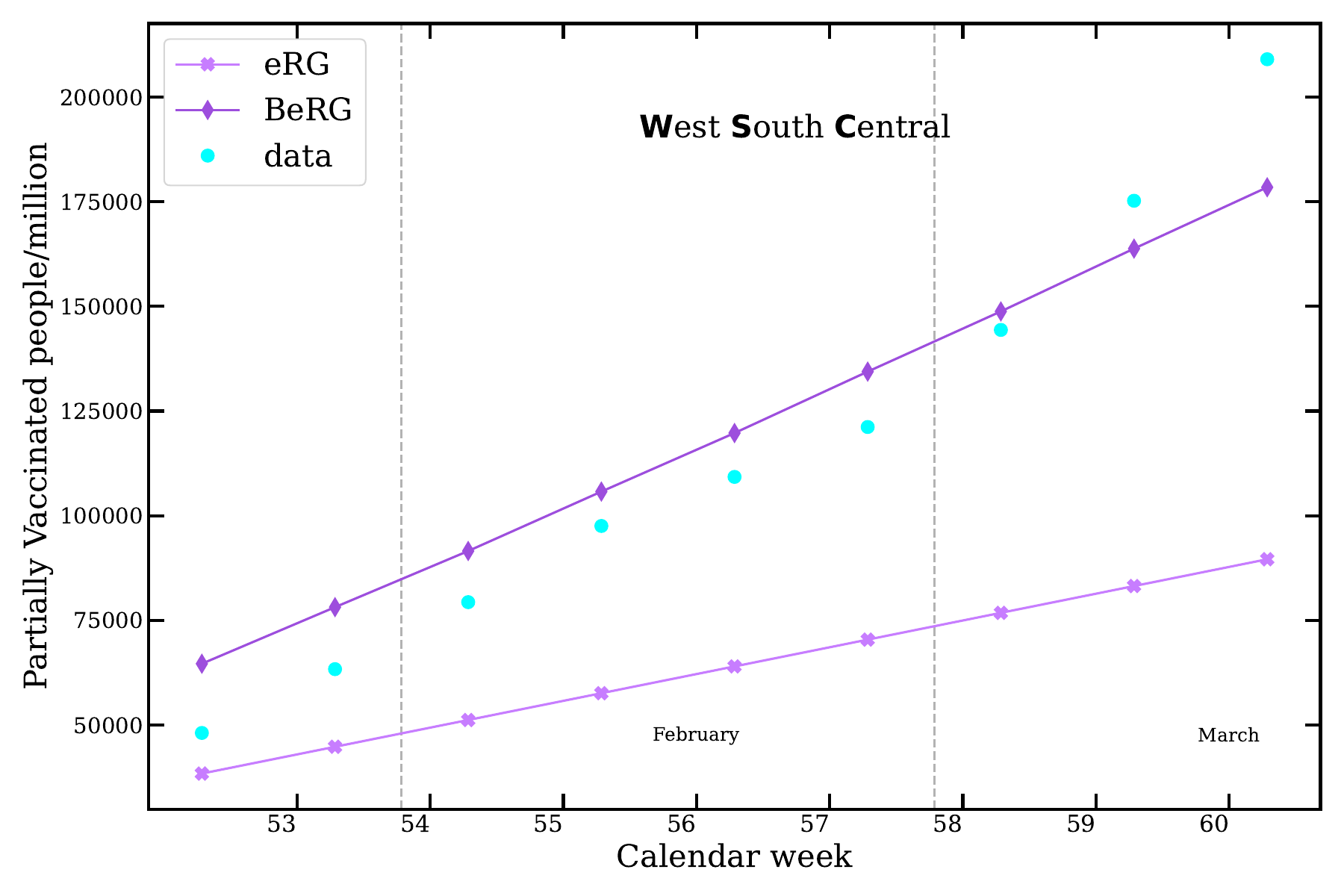}
		 \caption{WSC}
		 \label{fig:WSC}
	      \end{subfigure}
	     \begin{subfigure}{0.30\linewidth}
		 \includegraphics[width=\linewidth]{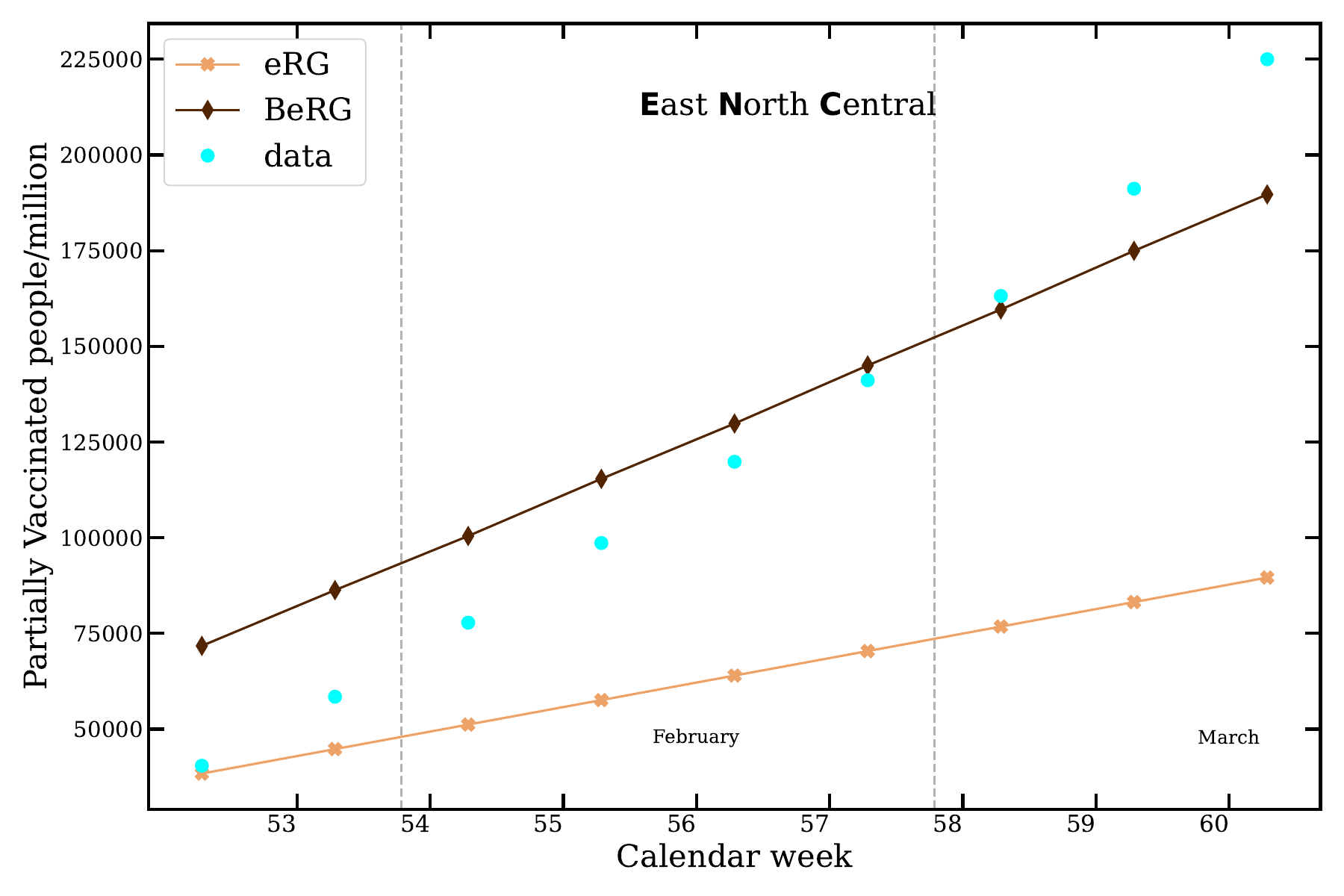}
		 \caption{ENC}
		 \label{fig:ENC}
	      \end{subfigure}
               \vfill
        \medskip
            \begin{subfigure}{0.30\linewidth}
    		  \includegraphics[width=\linewidth]{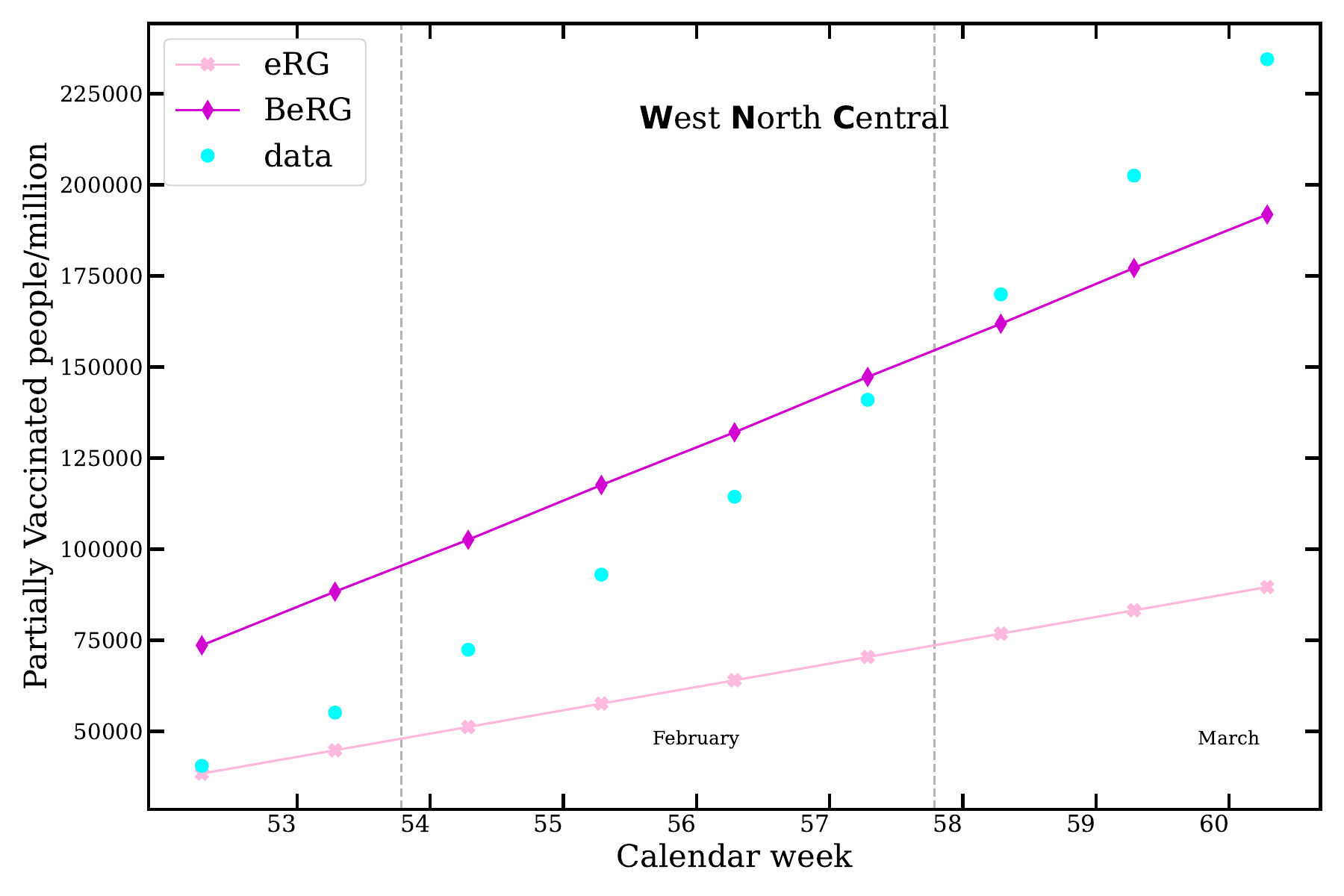}
    		  \caption{WNC}
    		  \label{fig:WNC}
    	       \end{subfigure}
    	     \begin{subfigure}{0.30\linewidth}
    		 \includegraphics[width=\linewidth]{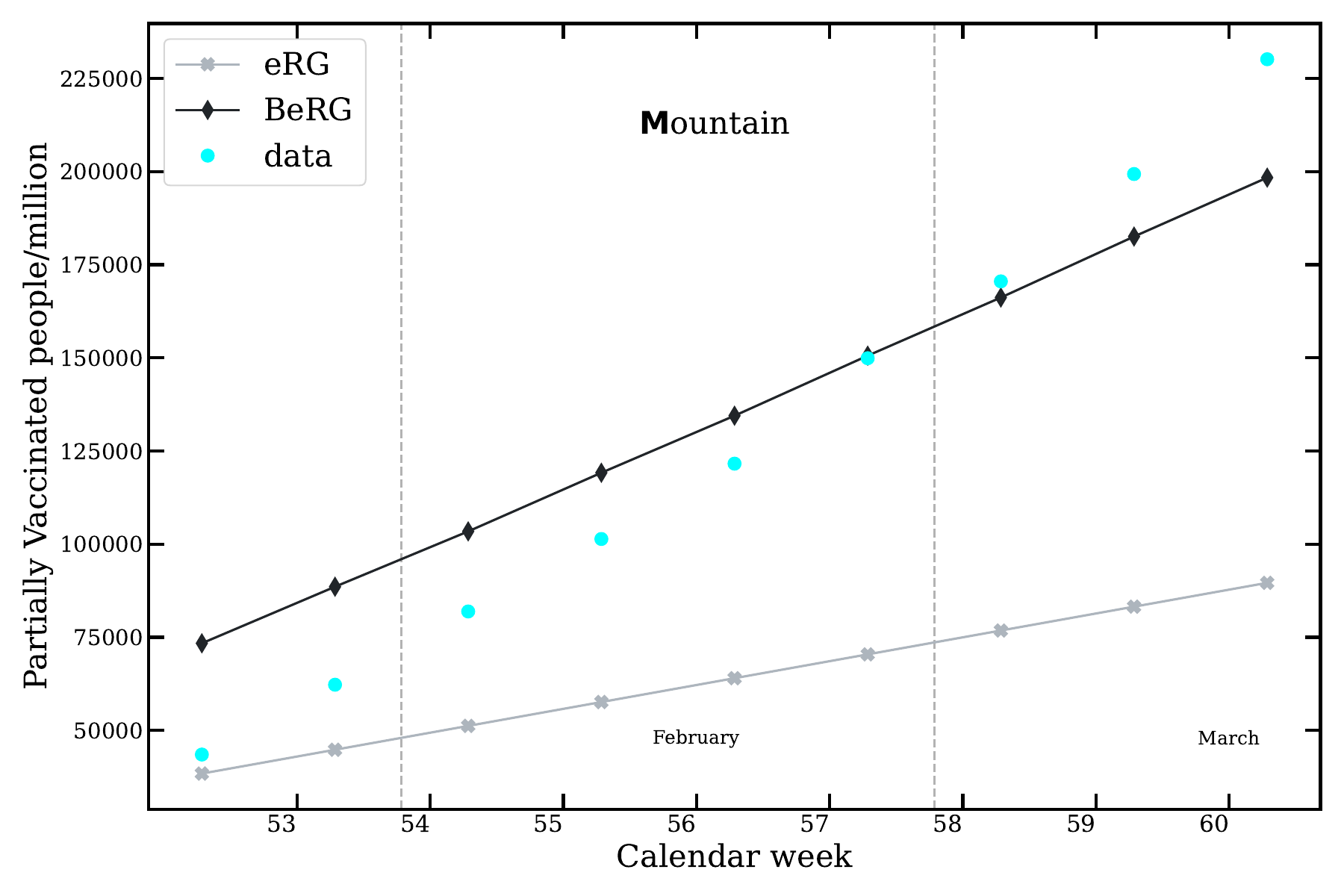}
    		 \caption{M}
    		 \label{fig:M}
    	      \end{subfigure}
    	     \begin{subfigure}{0.30\linewidth}
    		 \includegraphics[width=\linewidth]{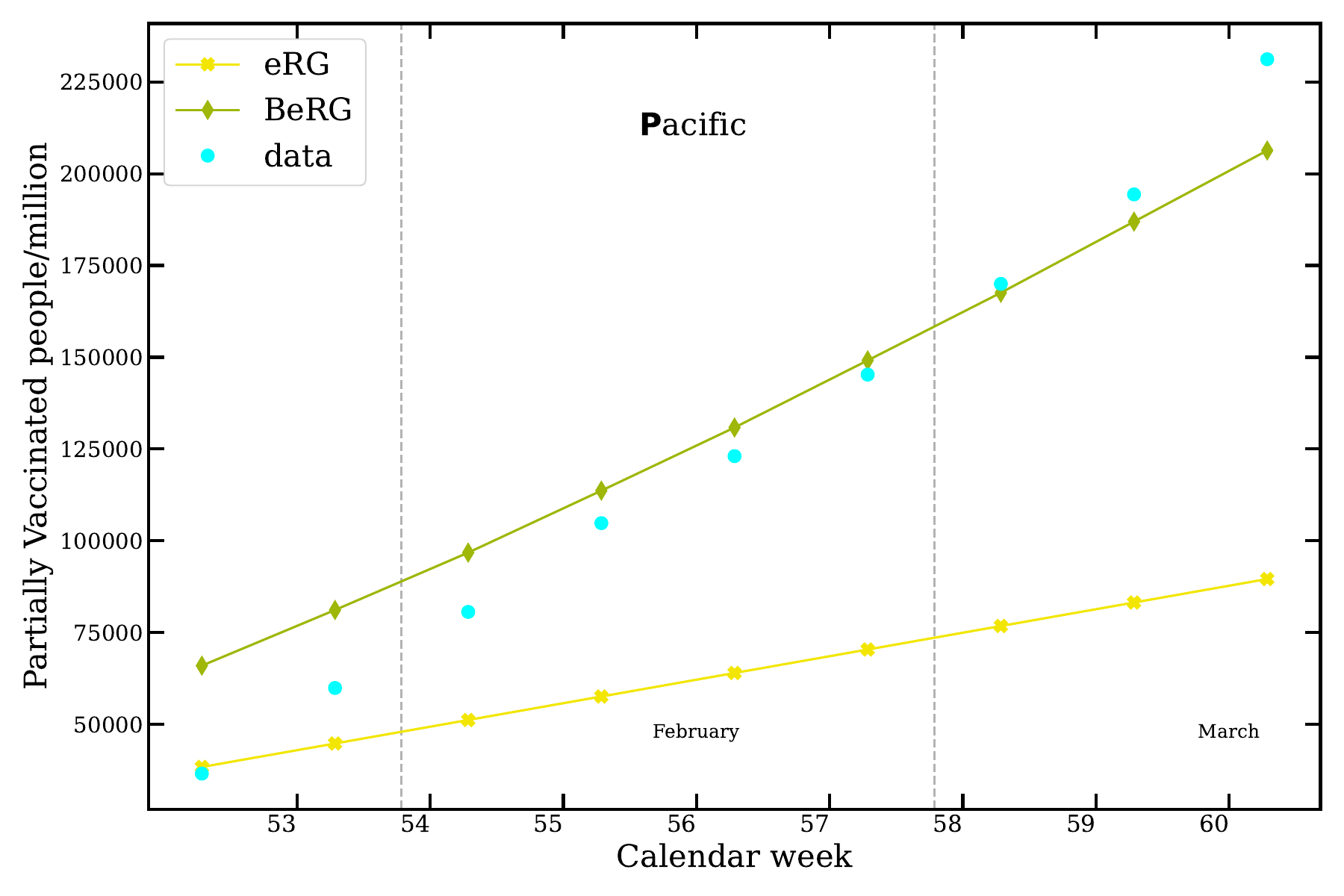}
    		 \caption{P}
    		 \label{fig:P}
    	      \end{subfigure}
    \caption{Plots of the official data of partial vaccinated people (in millions) $V_0$ and $V_{BeRG}$ together with the eRG predictions.}
    \label{fig:plotsPartVax}
\end{figure}

We are now ready to estimate the information coverage parameter $k$. This will be achieved  by demanding that the number of partially vaccinated individuals (per millions) from the model, during the vaccination campaign, is as much as possible close to the official number of partially vaccinated people $V_0$. Precisely, by \emph{partially vaccinated} we mean those individuals who have received the first dose of Pfizer or Moderna vaccines. The estimate is obtained minimising  
\begin{equation}
    \chi^2_i=\dfrac{1}{t_f-t_1-1}\sum_{l=1}^{f}\left(V_{BeRG, i}^l-V_{0, i}^l\right)^2\qq{with} i=1,\dots 9 
\end{equation}
with respect to $k$ for each division $i$, where
\begin{equation}
    V_{BeRG, i}^l=\int_{t_0}^{t_l}\ c(M_i) N_i d t\ ,
\end{equation}
is the number of vaccinated people from the BeRG model, $N_i$ is the population corresponding to the ith-division, $t_0$ marks the official start of the US vaccination campaign (i.e., December 16, 2020) and $t_f$ the date when we stop our simulation. The label $l$ increments the number of weeks at which we evaluate the $\chi^2_i$ with its  maximum value $f$ which for the first dose deployment corresponds to $f=9$.  For each division $i$, we computed $V_{BeRG, i}^l$ from week $t_1=26$ (corresponding to January 21, 2021) to week $t_9=34$ (March 18, 2021). We started the analysis at the end of January to minimize the effects due to the end of the year holidays and the one available from the eRG analysis in \cite{cot2021impact}.

The obtained values of the information coverage parameter $k$ for each US division are reported in Table \ref{tab:1} and graphically shown in Fig. \ref{fig:listak}. It ranges from a minimum of $k=0.47$ in the ESC division, to a maximum of $k=0.91$ in the SA division. The variability in the information coverage parameter depends on several factors influencing how the population experiences the pandemics in different US divisions. These range from personal believes to social-economical status, and are all integrated in the overall value for each $k$. 

To assess the performance of the the BeRG model, we perform the best fits and show in Fig. \ref{fig:plotsPartVax} the partially vaccinated individuals per million compared with eRG curves and official data. Since the procedure is similar for all the divisions, we elucidate here only the New England (NE) case.  First, we determine the value $c_0$ in \eqref{c0+c1} as $c_0=0.0064~{\rm week}^{-1}$,  in agreement with the official data regarding the initial phase of the vaccination campaign \cite{cot2021impact}. Then, the eRG curve  is reproduced by taking $c=c_0$ while the BeRG curve  is obtained by assuming  $c = c_0 + c_1(M)$, with $c_1(M)$ that further depends on the parameter $k$, previously estimated, via the the information index $M$. From Fig. \ref{fig:plotsPartVax} (a) it can be seen that for the division NE the BeRG model (blue line) is able to reproduce the official data (light blue dots) better than eRG (light blue line), as expected. In Figure \ref{fig:plotsPartVax} the same analysis is shown, with similar conclusions, for all the other US divisions. The improvement stemming from the BeRG versus the eRG can be quantitatively determined through the relative $\chi$-squared, as reported in Table~\ref{tab:chisquared}.

\begin{table}[t!]
\centering
\begin{tabular}{| c | c|}
\hline
\rowcolor{lightgray} \multicolumn{2}{|c|}{\textbf{eRG vs BeRG }}          \\ 
\rowcolor{lightgray} \multicolumn{2}{|c|}{\textbf{Vaccinated Individuals}}          \\ 
\hline \rowcolor{lightgray}
\textbf{Code}             &  $\mathbf{\dfrac{\chi^2_{eRG}-\chi^2_{BeRG}}{\chi^2_{eRG}}}$   \\ \cline{1-2} 
NE  & $0.92$                    
\\ \hline
MA  & $0.90$                 
\\ \hline
SA  & $0.96$                
\\ \hline
ESC & $0.91$              
\\ \hline
WSC & $0.88$                
\\ \hline
ENC & $0.90$                  
\\ \hline
WNC & $0.86$               
\\ \hline
M   & $0.92$              
\\ \hline
P   & $0.95$             
\\ 
   \hline
\end{tabular}
\caption{Relative error improvement in percentage between the BeRG model and eRG one with $c_0=c=0.0064$ computed starting from January 21, 2021  to March 18, 2021.}
\label{tab:chisquared}
\end{table}

\vspace{0.5cm}

\noindent {\it (c) Testing the validity of the BeRG}

\vspace{0.5cm}


\begin{table}[t!]
    \centering
    \begin{tabular}{|l|l|l|c|c|c|}
\hline \rowcolor{lightgray} \multicolumn{6}{|c|}{ \textbf{Second wave parameters}} \\
\hline \rowcolor{lightgray} \textbf{Division} & \textbf{Code} & $\mathbf{\log{A}}$ & $\mathbf{\gamma}$ & $\mathbf{k}$ & $\mathbf{D}$\\
\hline New England & NE & 11.036 & 0.204 & 0.52 & $5\times 10^{-6}$\\
\hline Mid-Atlantic & MA  & 11.075 & 0.185 & 0.75 & $8\times 10^{-6}$\\
\hline South Atlantic & SA  & 11.075 & 0.185 & 0.60 & $10^{-5}$\\
\hline East South Central & ESC & 11.221 & 0.207 & 0.64 & $5\times 10^{-6}$\\
\hline
\end{tabular}
    \caption{Parameters of the BeRG model for the second wave in 4 divisions: NE, MA, SA and ESC. The values are chosen to reproduce the data until March 4, 2021.}
    \label{tab:totale}
\end{table}

\begin{figure}[t!]
      \centering
	   \begin{subfigure}{0.45\linewidth}
		\includegraphics[width=\linewidth]{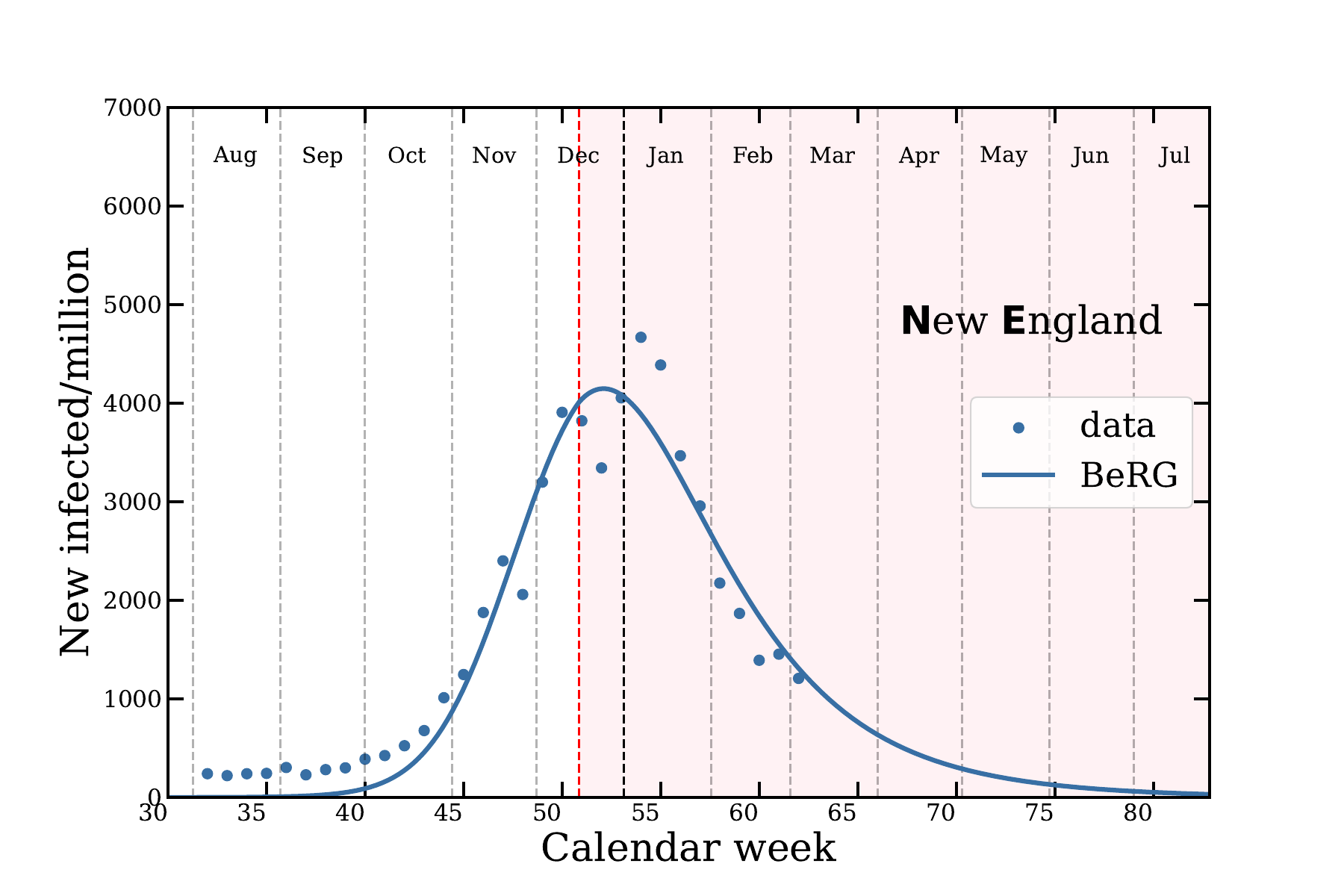}
		\caption{NE}
		\label{fig:NEtot}
	   \end{subfigure}
	   \begin{subfigure}{0.45\linewidth}
		\includegraphics[width=\linewidth]{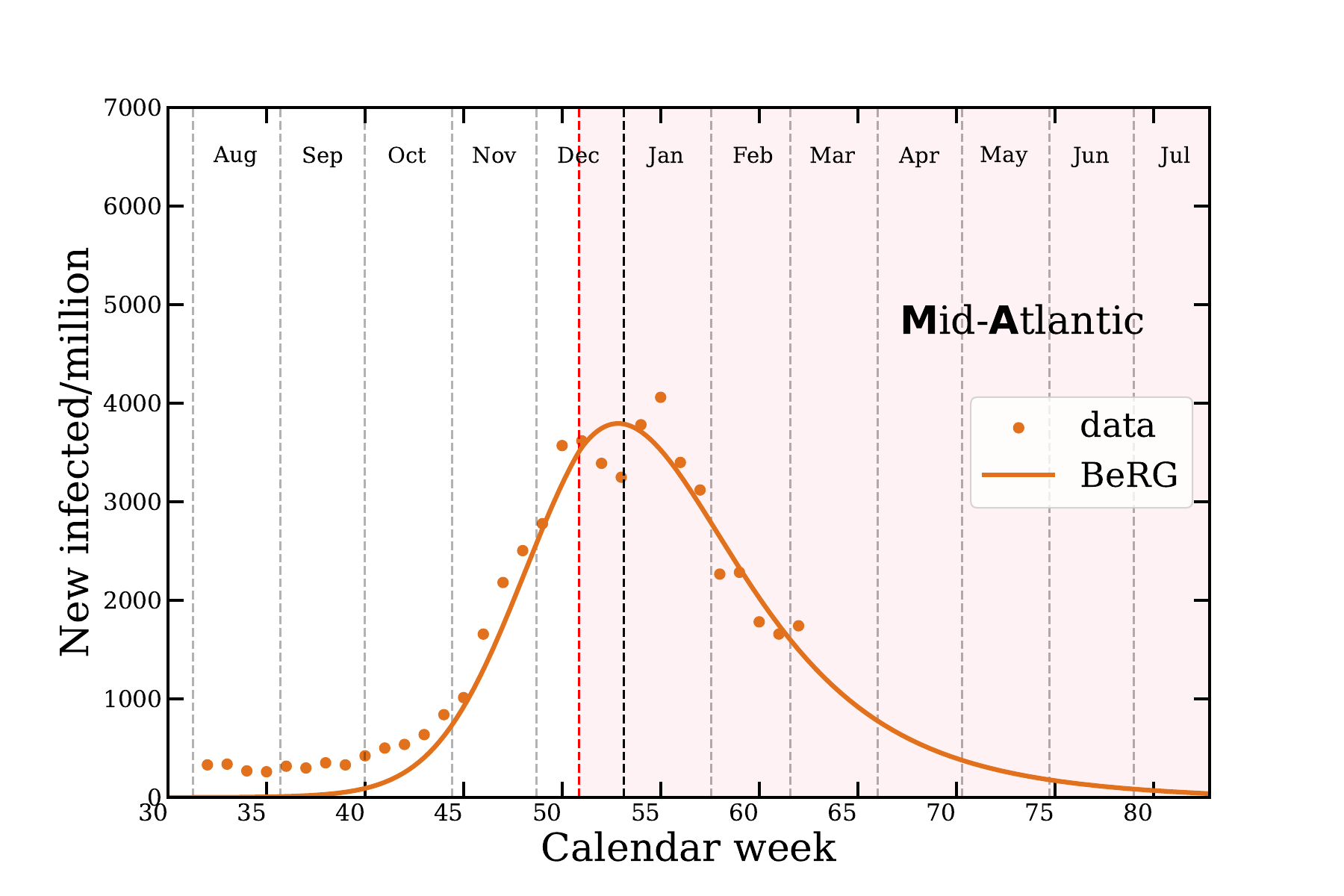}
		\caption{MA}
		\label{fig:MAtot}
	    \end{subfigure}\vfill
     \begin{subfigure}{0.45\linewidth}
		 \includegraphics[width=\linewidth]{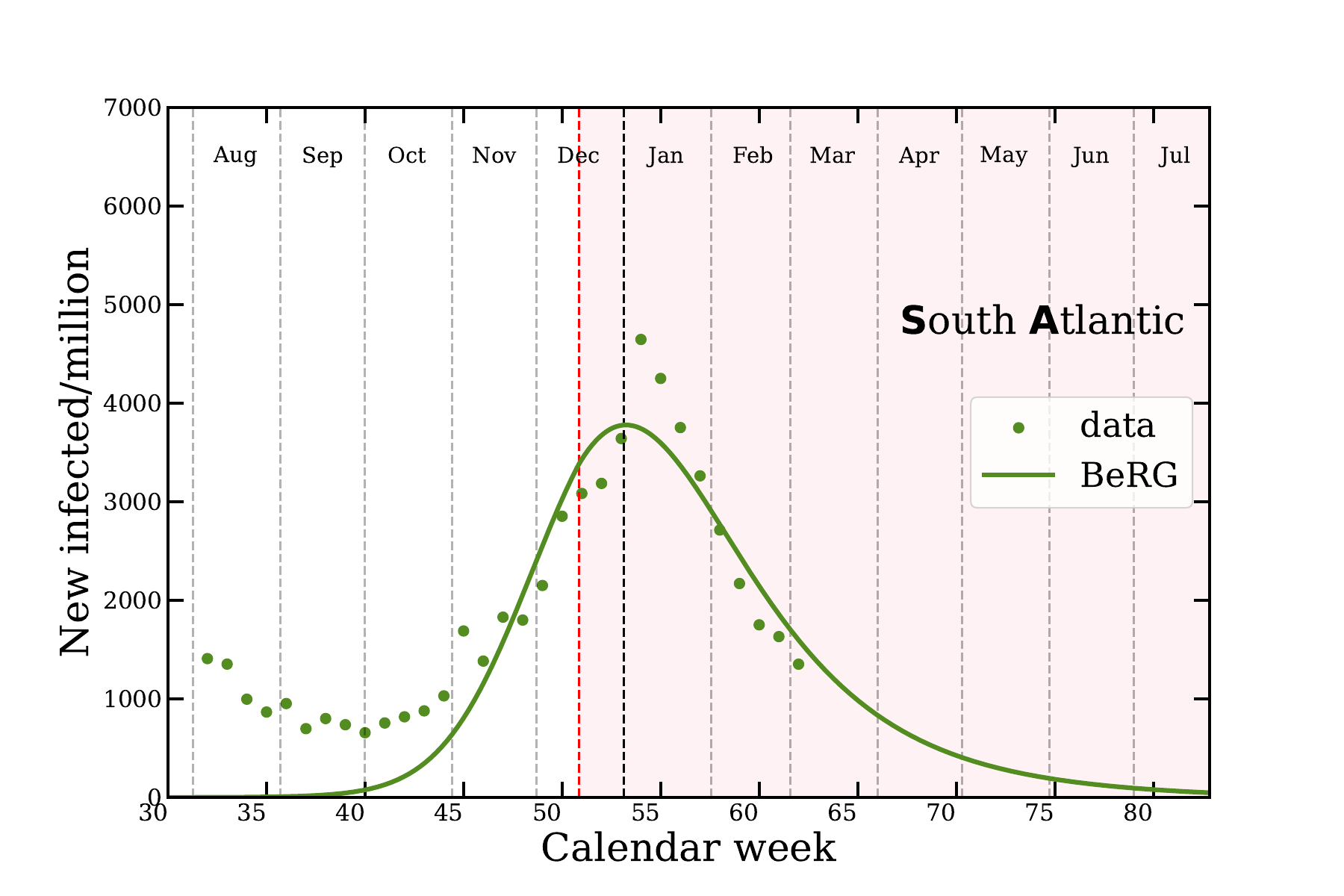}
		 \caption{SA}
		 \label{fig:SAtot}
	      \end{subfigure}
	       \begin{subfigure}{0.45\linewidth}
		  \includegraphics[width=\linewidth]{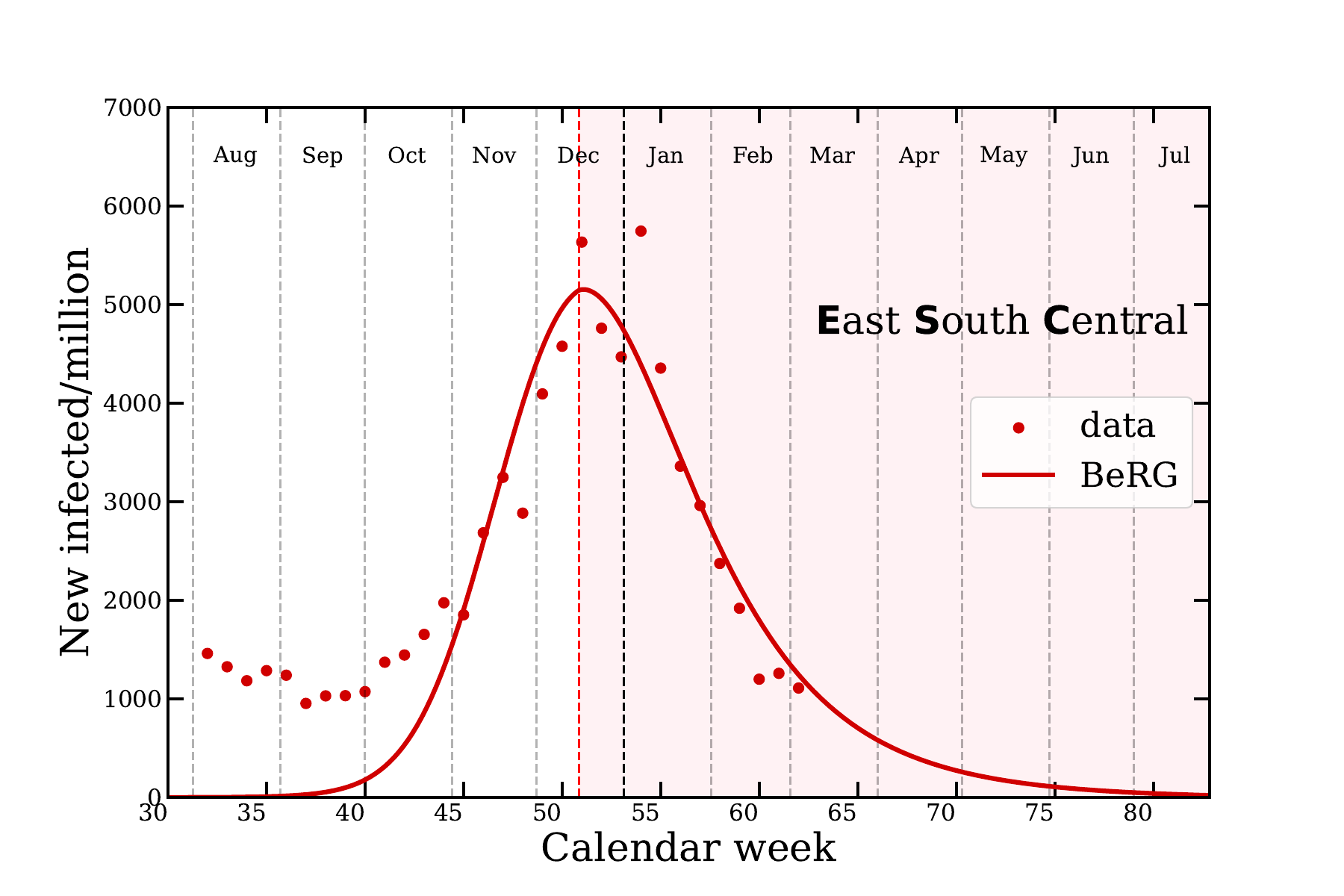}
		  \caption{ESC}
		  \label{fig:ESCtot}
	       \end{subfigure}
        \caption{Plots of the new infected/millions for the four divisions having the peak after the beginning of the vaccination compaign. The parameters of the BeRG model are listed in Table \ref{tab:totale}.}
        \label{fig:infettitotali}
\end{figure}

We now move to reconsider the pandemic evolution using the BeRG for those divisions where the vaccination campaign starts when the epidemic peak has not yet been achieved. In fact, if the peak has been already reached the BeRG has no chance to improve over the eRG since in this case the epidemic is already winding down and therefore the impact of the vaccination behaviour is curbed. The list of the divisions analyzed are reported in Table~\ref{tab:totale} alongside the newly determined values for $A$, the infection rate $\gamma$, the information coverage $k$ and the reactivity parameter $D$. The values of the information coverage $k$, taking into account its meaning defined in section 2,  describes the level of awareness of the population. For an example, for the East South Central region it was approximately 64$\%$ over the cumulative number of infectious individuals.  Additionally, the values of $k$ are consistent with those obtained in other studies,  based on field data, that adopt the information index approach \cite{BBRDMCovid,BBRDMmeningite}. Overall, we observe a high degree of similarity for the various parameters across the different divisions.

We report the final fits for the four regions Fig.~\ref{fig:infettitotali} showing a good understanding of the data via the BeRG, except for the high variability around the winter holidays whose description is beyond the scope of our analysis.

\section{Conclusions and outlook}
\label{sec:conclusions}

We introduced the information index augmented eRG model to describe vaccination behaviours during an epidemic outbreak. As a relevant test of the model we considered the US vaccination campaign for the COVID-19 epidemic. We performed multiple tests of the model by comparing it to the data corresponding to seven US divisions. We discover that, when comparing to the study in \cite{cot2021impact} without the behavioural component, the model provides a better representation of the data for four of the seven divisions. These are the divisions where the peak of the number of new infected is reached after the start of the vaccination campaign. The result is consistent with the expectation that the vaccination campaign and associated human behaviour has an impact when the epidemic is still rising. For the remaining three divisions the vaccination campaign starts around or after the peak of new infected leaving little space for the human behaviour to be relevant. Nevertheless, we still observe a strong behavioural impact on the increase of the number of vaccinated for all divisions. In fact, more people get vaccinated since the strength of the behavioural effect is modeled to increase with the cumulative number of infected. Last but not least the information parameter values estimated for this work are in line with the ones in related literature \cite{BBRDMCovid,BBRDMmeningite}.  The model can be readily applied to other regions of the world and could also include extra information stemming from  population studies by means of extensive surveys. 

\paragraph*{Acknowledgements}  
This work has been performed under the auspices of the
Italian National Group for Mathematical Physics (GNFM) of the
National Institute for Advanced Mathematics (INdAM).
This research was supported by EU funding within the NextGenerationEU---MUR PNRR Extended Partnership initiative on Emerging Infectious Diseases (Project no. PE00000007, INF-ACT).
B.B. also acknowledges PRIN 2020 project (No. 2020JLWP23) ``Integrated Mathematical Approaches to Socio -- Epidemiological Dynamics''.  A.D.A. expresses sincere appreciation to the University of Southern Denmark and D-IAS for their hospitality during the crucial stages of the work. The work of F.S. is
partially supported by the Carlsberg Foundation, semper ardens grant CF22-0922.
\appendix

\section{Beta function from traditional SIR model}\label{App_beta}

Let us consider an infectious disease spreading in a host population.
We assume that the epidemic dynamics is ruled by a classical SIR model but with time--varying coefficients:
\begin{equation}
    \begin{aligned}
&\dfrac{dS}{dt}=-\rho SI  , \\
&\dfrac{dI}{dt}=\rho S I  -\epsilon I , \\
&\dfrac{dR}{dt}=\epsilon I  ,\end{aligned}\label{ReducedSIRclassical}
\end{equation}
where $S$, $I$ and $R$ are, respectively, the fractions of susceptible, infectious and recovered individuals  with respect to the total population $N$ at time $t$; $\rho$ is the disease transmission rate at time $t$; $\epsilon$ is the recovery rate at time $t$.  
To model (\ref{ReducedSIRclassical}) we associate the following initial conditions:
\begin{equation}
    \label{IC}
S (0)=S _0>0,\,I (0)=I _0>0,\,R (0)=0,\end{equation}
with $S_0+I_0=1$.

By adding the three equations in (\ref{ReducedSIRclassical}), we obtain that the total population is constant: $S +I +R=1$ for all $t\geq 0$. Thus, since the dynamics of $R$ can be obtained as $R =1-S -I $, we limit to study  the differential equations for $S $ and $I $. 
Dividing the latter by the former, we get a differential
equation for $I$ as a function of $S$:
\begin{equation}
    \label{SIRparametric}
\dfrac{dI}{dS}=-1+\dfrac{1}{\sigma S},
\end{equation}
where we have introduced the ratio 
$$\sigma=\dfrac{\rho}{\epsilon},$$
which is a time--varying quantity. In particular, we assume  \textit{a priori} that $\sigma$
is  a function of $S$: 
$$\sigma=\sigma(S):[0,1]\to\mathrm{R}^+.$$ 
Following \cite{CacciapagliaABC} we  integrate the equation \eqref{SIRparametric} in the following form
$$
I(S)=1-S+\int_{S_0}^{S}\dfrac{du}{u\,\sigma(u)}\,,
$$
which is compatible with the initial conditions (\ref{IC}). Inserting this relation into the first equation of \eqref{ReducedSIRclassical}  yields
\begin{equation}
\dfrac{dS}{dt}=-\rho  S \left[1-S+\int_{S_0}^{S}\dfrac{du}{u\,\sigma(u)}\right]\,.\label{ParametricSeq}
\end{equation}
Instead of the fraction of susceptible individuals, the previous equation can be rewritten in terms of the cumulative number of infectious individuals $\mathcal{I}$, that is defined as
$$\mathcal{I}(t)=N\left(I_0+\int_0^t \rho S(\theta)I(\theta)d\theta\right).$$ 
Thus, 
$$S=1-\dfrac{\mathcal{I}}{N},$$
and
\begin{equation}
\dfrac{d\mathcal{I}}{dt}=-N\dfrac{d{S}}{dt}=N \rho \left(1-\dfrac{\mathcal{I}}{N}\right)\left[\dfrac{\mathcal{I}}{N}+\int_{S_0}^{1-{\mathcal{I}}/{N}}\dfrac{du}{u \sigma(u)}\right].\label{IcEq}
\end{equation}
Finally, by substituting the equation \eqref{IcEq} in the general formulation of the $\beta$ function (\ref{eq:betap}), we obtain the following expression directly derived from the epidemic model (\ref{SIRparametric}):
\begin{equation}
-\beta(\alpha)=\dfrac{d\alpha}{d\mathcal{I}}N\rho\left(1-\dfrac{\mathcal{I}}{N}\right)\left[\dfrac{\mathcal{I}}{N}+\int_{S_0}^{1-{\mathcal{I}}/{N}}\dfrac{du}{u\sigma(u)}\right].\label{FormBetaFromSIR}
\end{equation} 

\printbibliography
\end{document}